\documentclass[twocolumn,showpacs,preprintnumbers,amsmath,amssymb,superscriptaddress]{revtex4}

\usepackage{graphicx}
\begin{document}
\bibliographystyle{prsty}
\title{Bulk-like $d$ band of SrVO$_{3}$ under a self-protective cap layer}

\author{M.~Takizawa}
\affiliation{Department of Physics, University of Tokyo, 
Bunkyo-ku, Tokyo 113-0033, Japan }
\author{M.~Minohara}
\affiliation{Graduate School of Arts and Sciences, University of Tokyo, 3-8-1 Komaba, Meguro-ku, Tokyo 153-8902, Japan}
\author{H.~Kumigashira}
\affiliation{Department of Applied Chemistry, University of Tokyo, 
Bunkyo-ku, Tokyo 113-8656, Japan}
\affiliation{Research for Evolutional Science and Technology of Japan Science and Technology Corporation (JSTCREST), Chiyoda-ku, Tokyo 102－0075, Japan}
\author{D.~Toyota}
\affiliation{Department of Applied Chemistry, University of Tokyo, 
Bunkyo-ku, Tokyo 113-8656, Japan}
\author{M.~Oshima}
\affiliation{Graduate School of Arts and Sciences, University of Tokyo, 3-8-1 Komaba, Meguro-ku, Tokyo 153-8902, Japan}
\affiliation{Department of Applied Chemistry, University of Tokyo, 
Bunkyo-ku, Tokyo 113-8656, Japan}
\affiliation{Research for Evolutional Science and Technology of Japan Science and Technology Corporation (JSTCREST), Chiyoda-ku, Tokyo 102－0075, Japan}
\author{H.~Wadati}
\affiliation{Department of Physics, University of Tokyo, 
Bunkyo-ku, Tokyo 113-0033, Japan }
\author{T.~Yoshida}
\affiliation{Department of Physics, University of Tokyo, 
Bunkyo-ku, Tokyo 113-0033, Japan }
\author{A.~Fujimori}
\affiliation{Department of Physics, University of Tokyo, 
Bunkyo-ku, Tokyo 113-0033, Japan }
\author{M.~Lippmaa}
\affiliation{Institute for Solid State Physics, University of Tokyo, 5-1-5 Kashiwanoha, Kashiwashi, Chiba, 277-8581, Japan}
\author{M.~Kawasaki}
\affiliation{Institute for Materials Research, Tohoku University, 2-1-1 Katahira, Aoba-ku, Sendai, Miyagi, 980-8577, Japan}
\affiliation{Research for Evolutional Science and Technology of Japan Science and Technology Corporation (JSTCREST), Chiyoda-ku, Tokyo 102－0075, Japan}
\author{H.~Koinuma}
\affiliation{Materials and Structures Laboratory, Tokyo Institute of Technology, 4259 Nagatsuta, Midori-ku, Yokohama, Kanagawa, 226-8503, Japan}
\author{G.~Sordi}
\affiliation{Departamento de F\'isica, FCEN, Universidad de Buenos Aires, Ciudad Universitaria Pabell\'on I, (1428) Buenos Aires, Argentina}
\author{M.~Rozenberg}
\affiliation{Departamento de F\'isica, FCEN, Universidad de Buenos Aires, Ciudad Universitaria Pabell\'on I, (1428) Buenos Aires, Argentina}
\date{\today}

\begin{abstract}
We have performed a detailed angel-resolved photoemission spectroscopy study of {\it in-situ} prepared SrVO$_3$ thin films. 
Naturally capped by a ``transparent'' protective layer, contributions from surface states centered at $\sim -1.5$ eV are dramatically reduced, enabling us to study the bulk V $3d$ states. 
We have observed a clear band dispersion not only in the coherent quasiparticle part but also in the incoherent part, which are reproduced by dynamical mean-field theory calculations and the spectral weight of the incoherent part is stronger within the Fermi surface. 
\end{abstract}


\maketitle
Metal-insulator transition (MIT) has been extensively studied because of its fundamental importance in condensed matter physics \cite{MIT}. 
To study bandwidth-control MIT, perovskite-type oxides {\it AB}O$_3$ having formally the $d^1$ electronic configuration are ideal systems and have been studied extensively \cite{Fujimori-d1, CSVO-Inoue1, CSVOpes-Inoue}. 
In {\it AB}O$_{3}$, the bandwidth $W$ is controlled through the modification of the radius of the {\it A}-site ion, and hence the {\it B}-O-{\it B} bond angle $\theta$. 
On the theoretical side, the recent development of dynamical mean-field theory (DMFT) has led to a major progress in understanding MIT in strongly correlated systems \cite{DMFT, Sordi}. 
According to DMFT, as $U/W$ increases, where $U$ is the on-site Coulomb energy, spectral weight is transferred from the coherent part [the quasiparticle band near Fermi level ({\it E$_\mathrm{F}$})] to the incoherent part (the remnant of the Hubbard bands $\sim 1 - 2$ eV above and below {\it E$_\mathrm{F}$}) \cite{Zhang}. 

Ca$_{1-x}$Sr$_{x}$VO$_{3}$ (CSVO) is a typical bandwidth-controlled system, where the average $A$-site ionic radius and hence the bandwidth $W$ can be continuously changed, although it remains metallic in the entire $x$ range \cite{CSVO-Inoue1}. 
Extensive studies on CSVO have been made to understand the evolution of the electronic structure as a function of $U/W$ but the result remains highly controversial till now. 
In an early photoemission spectroscopy (PES) study of CSVO, Inoue {\it et al.} \cite{CSVOpes-Inoue} have reported that as one decreases $x$ in CSVO, that is, as one increases $U/W$, spectral weight is transferred from the coherent part to the incoherent part centered $\sim$ 1.5 eV below {\it E$_\mathrm{F}$}. 
By measuring the photon-energy-dependent PES of CSVO, Maiti {\it et al.} \cite{CSVO-maiti} have pointed out that the surface states are different from the bulk ones. 
Sekiyama {\it et al.} \cite{CSVO-Sekiyama} have reported that the spectral weight of the incoherent part is considerably reduced and that spectral weight transfer is not observed in the PES spectra of CSVO taken using high energy photons, i.e., in so-called ``bulk-sensitive" PES spectra. 
Another ``bulk-sensitive" PES spectra using very low energy photons \cite{CSVO-Eguchi}, however, has indicated small differences between SrVO$_3$ (SVO) and CaVO$_3$ (CVO) qualitatively consistent with Inoue {\it et al.} \cite{CSVOpes-Inoue}. 
Moreover, a PES study using polarized and unpolarized photons has indicated that matrix-element effect is important and that there is finite spectral weight transfer from the coherent part to the incoherent part in going from SVO to CVO \cite{CSVO-maiti2}. 
A recent extended cluster-model analysis of the PES spectra, on the other hand, has suggested that contributions from the O $2p$ states are more significant in the incoherent part than in the coherent part and that the reduction of spectral weight of the incoherent part at high photon energies is partly due to the effect of atomic-orbital cross-sections \cite{Mossanek}. 
In contrast to the controversial results mentioned above, a recent angle-resolved photoemission spectroscopy (ARPES) study by Yoshida $et$ $al$. \cite{SVO-yoshida} have given clear-cut information. 
They successfully performed an ARPES study of cleaved bulk SVO crystals and observed the band dispersions of the coherent part \cite{SVO-yoshida}. 
Recent DMFT calculations have given $\bf{k}$-resolved spectral functions and predicted a kink in the quasiparticle band dispersion due to the energy dependence of the self-energy \cite{Nekrasov, Byczuk}. 

In the present work, we have fabricated SVO thin films having atomically flat surfaces using the pulsed laser deposition (PLD) technique and studied its detailed electronic structure by {\it in-situ} ARPES measurements. 
ARPES experiments on thin films prepared {\it in-situ} has recently turned out to have many advantage because their well-defined atomically flat surface enable us to obtain more detailed ARPES spectra \cite{LSMO2-chikamatsu, PCMO-wadati}. 
Such high-quality samples dramatically reduce surface states, and a clear dispersion of intrinsic origin can be observed. 

The PES measurements were performed at BL-1C, 2C, and 28B of Photon Factory (PF), High Energy Accelerators Research Organization (KEK), using a combined laser molecular beam epitaxy-photoemission spectrometer system. 
Details of the experimental setup are described in Ref.~\cite{Horiba}. 
Epitaxial thin films of SVO were grown on single-crystal substrates of Nb-doped SrTiO$_3$ by the PLD method. 
The substrates were annealed at 1050~${}^{\circ}$C under an oxygen pressure of $\sim 1\times 10^{-6}$ Torr to obtain an atomically flat TiO$_2$-terminated surface \cite{kawasaki}. 
SVO thin films were deposited on the substrates at 900~${}^{\circ}$C under a high vacuum of $\sim 10^{-8}$ Torr. 
The surface morphology of the measured SVO thin films was checked by {\it ex-situ} atomic force microscopy, showing atomically flat step-and-terrace structures. 
The crystal structure was characterized by four-circle X-ray diffraction, and a coherent growth on the substrate was confirmed. 
The in-plane lattice constant was $a = 3.905$ \AA, the same as that of SrTiO$_3$, while the out-of-plane lattice constant was $c = 3.82$ \AA. 
All the PES measurements were performed in an ultrahigh vacuum of $\sim 10^{-10}$~Torr at 20~K (unless otherwise noted) using a Scienta SES-100 electron-energy analyzer. The total energy resolution was set to about 30, 150, and 150-400 meV for the spectra near {\it E$_\mathrm{F}$} (BL-28B), in the entire valence-band region (BL-1C), and the spectra measured using soft x-rays (BL-2C), respectively. The {\it E$_\mathrm{F}$} position was determined by measuring gold spectra. 

\begin{figure}
\includegraphics[width=.8\linewidth]{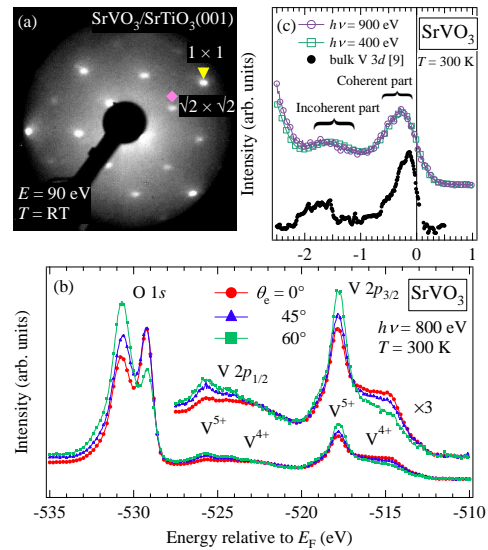}
\caption{(Color online) Surface characterization of SrVO$_3$ thin films synthesized by the PLD method. 
(a) LEED pattern showing $1 \times 1$ spots (indicated by triangle) as well as $\sqrt{2} \times \sqrt{2}$ surface super-structure spots (indicated by diamond). 
(b) O~$1s$ and V~$2p$ core-level spectra for various emission angle $\theta_{\text{e}}$ relative to the surface normal. 
Both spectra show two components which can be assigned to bulk (SrVO$_3$) and surface (V$^{5+}$ oxide) components, according to the angular dependence. 
(c) V~$3d$ emission near {\it E$_\mathrm{F}$}. 
The line shape is independent of photon energies, i.e., independent of the probing depth, and is similar to the bulk one \cite{CSVO-Sekiyama}. }
\label{character}
\end{figure}
Results of the characterization of the surfaces of the SVO films thus prepared are shown in Fig.~\ref{character}. 
As in Fig.~\ref{character}~(a), low energy electron diffraction (LEED) patterns show sharp $1 \times 1$ spots, with some super-structure spots of $\sqrt{2} \times \sqrt{2}$. 
For the O~$1s$ core level shown in Fig.~\ref{character}~(b), one can see two components. 
Because we performed {\it in-situ} PES measurements, the strong high binding energy structure of O~$1s$ cannot be attributed to contaminations. 
For the V $2p$ core level in the same panel, too, there appear two components which can be assigned to V$^{4+}$ and V$^{5+}$ \cite{Sawatzky}. 
With increasing emission angle $\theta_{\text{e}}$, i.e., with increasing surface sensitivity, the higher binding energy structure of O~$1s$ and the V$^{5+}$ component increase. 
This indicates that the surface of the SVO thin films thus fabricated is covered with a ``V$^{5+}$ oxide'' having a $\sqrt{2} \times \sqrt{2}$ structure. 
From the emission-angle dependence of the intensity of the O~$1s$ and V~$2p$ emission, the thickness of the surface ``V$^{5+}$ oxide'' layer is estimated to be 2 - 4 \AA, consistent with a 0.5 - 1 unit cell thickness of SVO. 
The structure of the surface V$^{5+}$ oxide layer is not known at present, but a possible candidate is a layer of one unit cell thickness with composition Sr$_{0.5}$VO$_3$ or SrVO$_{3.5}$, where Sr vacancies or excess oxygen atoms form a $\sqrt{2} \times \sqrt{2}$ super-lattice and V is oxidized to V$^{5+}$. 

As shown in Fig.~\ref{character}~(c), the V $3d$ spectra taken at 400 and 900 eV are identical to each other in spite of the different surface sensitivities. 
The relative intensity of the coherent part to the incoherent part is very high, similar to that of the ``bulk" spectra reported in Refs. \cite{CSVO-Sekiyama, CSVO-maiti, CSVO-maiti2}. 
This fact indicates that contributions from surface states seen in the previous work \cite{CSVO-maiti, CSVO-Sekiyama} were strongly reduced in the present SVO thin films. 
Due to the suppression of the surface states, the V $3d$ bulk electronic structures could be successfully investigated in APRES spectra, as we shall see below. 
Considering the observation of the $\sqrt{2} \times \sqrt{2}$ surface reconstruction and the V$^{5+}$ component in the V $2p$ core-level spectra as shown in Fig.~\ref{character}, we infer that the V$^{5+}$-oxide capping layer has no electronic states within $\sim 2$ eV of {\it E$_\mathrm{F}$}, protects the surface of SVO and hence the V $3d$ bands survive as in the bulk states within the band gap of the V$^{5+}$ ($d^0$) oxide surface layer, just like in the case of the SrTiO$_3$ capped LaTiO$_3$ \cite{STOLTOtakizawa}. 

\begin{figure}
\includegraphics[width=\linewidth]{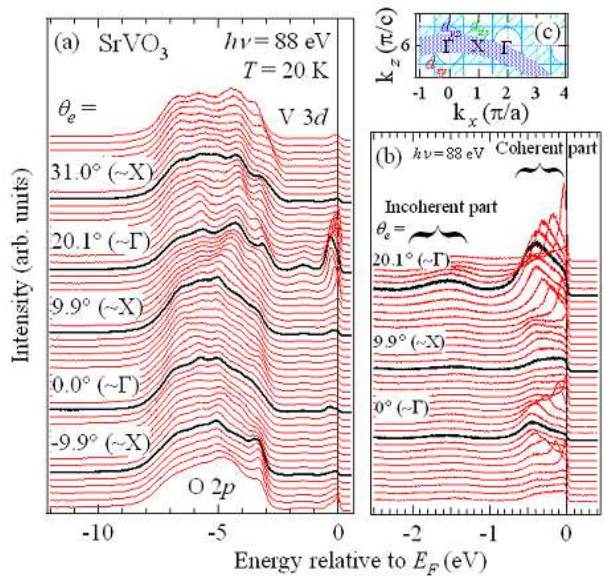}
\caption{(Color online) EDCs of SrVO$_3$ thin films. (a) Entire valence-band region. (b) Near the Fermi level. (c) Traces in $k$-space including the uncertainties in the $k_z$ direction. Schematic Fermi surfaces of SrVO$_3$ (junglegym-type mutually penetrating cylinders) are indicated by shaded area. }
\label{VB}
\end{figure}
A series of ARPES spectra or energy distribution curves (EDCs) taken at 88 eV are shown in Fig.~\ref{VB}~(a) and (b) for various emission angles $\theta_{\text{e}}$. 
The corresponding trace in $k$-space is shown in Fig.~\ref{VB}~(c). 
Here, the free-electron final states, the work function of the sample $\phi = 4.5$ eV, and the inner potential $V_0 = 11$ eV have been assumed to obtain the momentum $k_z$ perpendicular to the surface. 
Although $k_z$ is dependent on $\theta_{\text{e}}$ to some extent as shown in Fig.~\ref{VB}~(c), the ARPES spectra taken at 88 eV can be considered to trace the band in the $\Gamma$ - X direction since band dispersions along high symmetry lines are emphasized due to the finite width of $k_z$ and the high $k_x, k_y$-resolved density of states around the high symmetry lines \cite{TB-wadati}. 
For simplicity, we have also assumed the cubic Brillouin zone although the unit cell of SVO thin films is slightly distorted to the tetragonal structure ($c/a \simeq 0.978$). 
Figure~\ref{VB}~(a) shows clear dispersions both in the O $2p$ and V $3d$ bands. 
The bands between $-10$ eV to $-3$ eV are mainly composed of O $2p$ states while those between $-3$ eV to {\it E$_\mathrm{F}$} mainly of V $3d$ states. 
In the V $3d$ band [Fig.~\ref{VB}~(b)], there are two features; the coherent part (the quasiparticle band near {\it E$_\mathrm{F}$}) and the incoherent part (regarded as the remnant of the lower Hubbard band). 
As reported previously \cite{SVO-yoshida}, the coherent part shows clear dispersion. 
It should be noted that the coherent peak becomes much sharper and more intense than in the previous work \cite{SVO-yoshida} due to the well-defined atomically flat surfaces of thin films and probably the protective V$^{5+}$ oxide layer. 
As for the incoherent part, there is a weak but finite dispersion as we shall discuss below in detail. 

\begin{figure}
\begin{center}
\includegraphics[width=.8\linewidth]{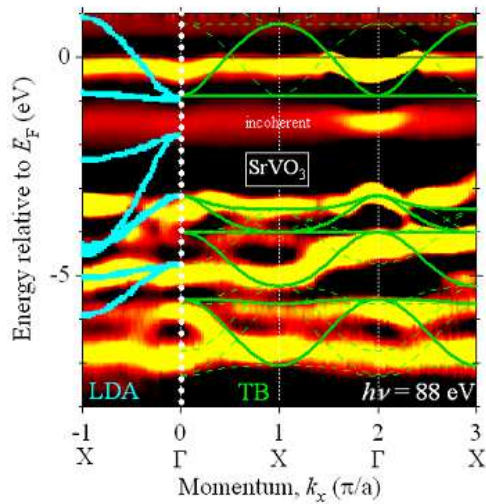}
\caption{(Color online) Experimental band structure deduced from the second derivative of EDCs, with bright parts corresponding to energy bands. LDA calculations from Ref.~\cite{takegahara} and the fitted tight-binding (TB) bands are also shown in the momentum range from $k_x = -\pi/a$ to 0 and from $k_x = 0$ to $3\pi/a$, respectively. As for the TB bands, the diffraction replica with respect to the $\sqrt{2} \times \sqrt{2}$ surface super-structure are also shown as broken curves. }
\label{band}
\end{center}
\end{figure}

In order to see the band dispersions more clearly, we have taken the second derivatives of the EDCs and displayed as grey-scale plot in Fig.~\ref{band}. 
(In this figure, bright parts correspond to energy bands.) 
Although the obtained band dispersions qualitatively agree with LDA calculation (left-hand side of Fig.~\ref{band}) \cite{Nekrasov, takegahara, Pavarini-d1}, the position of the O $2p$ bands does not. 
Therefore, we have performed tight-binding band calculation to adjust the O~$2p$ band position \cite{TB-wadati}. 
As shown in Fig.~\ref{band}, one obtains good agreement between experiment and calculation when we set the parameters: $\epsilon_d - \epsilon_p = 4.2$ eV, $\epsilon_{d\sigma} - \epsilon_{d\pi} = 2.3$ eV, $\epsilon_{p\sigma} - \epsilon_{p\pi} = 0.7$ eV, $(pd\sigma) = -2.2$ eV, $(pd\pi) = 1.4$ eV, $(pp\sigma) = 0.4$ eV, and $(pp\pi) = 0$ eV. 
This was also confirmed by comparing the calculation with the other band dispersion along the X - M line obtained using $h\nu = 60$ eV (not shown). 
Some of the bands, especially in the O~$2p$ band region, which cannot be reproduced by the calculation may be attributed to the presence of the V$^{5+}$ oxide surface layer. 
The detailed electronic structure of the V$^{5+}$ oxide surface layer is not clear now, however, taking into account of the diffraction replica with respect to the $\sqrt{2} \times \sqrt{2}$ surface super-structure, almost all the experimental O~$2p$ bands are well reproduced by the calculation (Fig.~\ref{band}). 

\begin{figure}
\begin{center}
\includegraphics[width=\linewidth]{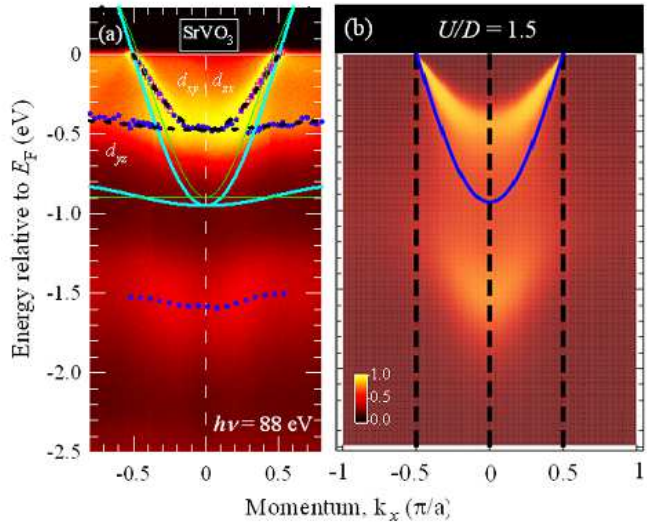}
\caption{(Color online) Energy- and momentum-dependent spectral weight near the Fermi level. (a) Experimental intensity plot for SrVO$_3$. Peak positions of the EDCs and MDCs are shown by filled circles and open squares, repectively. The V $3d$ bands from the LDA calculation \cite{LDA-Pavarini} and tight-binding calculation are also shown by solid thick and thin curves, respectively. Broken curves are LDA bands renormalized by a factor of 2. (b) Intensity plot of spectral function from DMFT calculation with $U/D = 1.5$. }
\label{NF}
\end{center}
\end{figure}

In order to see the V $3d$ bands near {\it E$_\mathrm{F}$} in more detail, we show in Fig.~\ref{NF}~(a) the $E$ - $k_x$ spece intensity plot along the $\Gamma$ - X direction \cite{intplot}. 
The peak positions determined from both EDCs and momentum distribution curves (MDCs) are also shown. 
The V $3d_{xy}$ and $3d_{zx}$ bands cross the Fermi level between the $\Gamma$ and X points. 
For the dispersion of the coherent part, one can see clear mass renormalization compared with the LDA calculation \cite{LDA-Pavarini}. 
From the experimental ($-0.44 \pm 0.02$ eV) and calculated ($-0.95$ eV) occupied bandwidths, the global mass renormalization factor is estimated to be $\sim 2$. 
That is, if the LDA band dispersions is reduced by a factor of 0.5, the experimental band dispersions are well reproduced as shown in Fig.~\ref{NF}~(a). 
This indicates that electron correlation strength is almost independent of momentum and of the $d_{xy}$, $d_{yz}$ or $d_{zx}$ bands of the degenerate $t_{2g}$ band. 
The kink in the band dispersion is weak and broad, if exists, but the curvature changes its sign around $\sim -0.2$ eV as predicted by a recent DMFT calculation \cite{Nekrasov}. 
As for the incoherent part located around $-1.5$ eV, one can see a weak but finite ($\sim 0.1$ eV) dispersion. 
The intensity of the incoherent part is momentum dependent and becomes strong within the Fermi surface. 

Figure~\ref{NF} (b) shows the intensity plot of the spectral functions from the DMFT calculation \cite{intplot}. 
The DMFT self-energy was computed using a single band model in the present case. 
One obtains agreement between experiment and theory when the correlation strength of $U/D$ is set to $1.5$, where $D$ is the bandwidth of the occupied part of the non-interacting band. 
Although the DMFT calculation predicts that an incoherent part disperses as strongly as the bare band, the experimental dispersion of the incoherent part was weaker. 
This is probably due to the overlapping dispersiveless $d_{yz}$ band along the $\Gamma$ - X direction, which has been neglected in the present DMFT calculation. 
In future, DMFT + LDA calculation which takes into account the three-fold degenerate of the $t_{2g}$ orbitals are necessary to quantitatively understand the ARPES results. 

In conclusion, we have studied the electronic structure of SrVO$_3$ thin films by means of ARPES. 
Due to the ``transparent'' protective surface V$^{5+}$ oxide layer, bulk-like V $3d$ band structure was successfully observed. 
We have determined the occupied quasiparticle width of the V $3d$ band to be $0.44 \pm 0.02$ eV. 
The band dispersions in the coherent part were reproduced by the renormalized LDA bands with the global mass renormalization factor of $\sim 2$. 
There was a weak but finite dispersion in the incoherent part and its intensity was stronger within the Fermi surface. 
The experimental dispersions and intensities of the coherent part as well as of the incoherent part were reproduced by momentum-resolved DMFT calculation. 
Since we have employed the single-band model for the DMFT calculation, multi-orbital effect of the $t_{2g}$ bands remains to be studied in future studies. 

The authors would like to thank K.~Ono and A.~Yagishita for their support in the experiment at KEK-PF and T.~Mizokawa and K.~Nasu for enlightening discussion. 
This work was supported by a Grant-in-Aid for Scientific Research (A19204037 and   A19684010) from JSPS and a Grant-in-Aid for Scientific Research in Priority Areas ``Invention of Anomalous Quantum Materials" from MEXT. 
Two of us (MT and HW) were supported by JSPS. 
The work was done under the approval of Photon Factory Program Advisory Committee (Proposal Nos. 2005G101 and 2005S2-002) at the Institute of Material Structure Science, KEK.


\begin{thebibliography}{10}
\bibitem{intplot}
In order to display the incoherent part visibly, the intensity has been modified using
  a hyperbolic tangent function after having appropriately normalized the experimental
  data.
\end{thebibliography}
\end{document}